\documentclass[preprint,10pt,authoryear]{elsarticle}
\usepackage{graphicx, verbatim, url}

\journal{JASIST}

\begin{document}
\begin{frontmatter}

\title{The large-scale structure of journal citation networks}

\author{Massimo Franceschet}

\address{Department of Mathematics and Computer Science, University of Udine \\
           Via delle Scienze 206 -- 33100 Udine, Italy \\
           \texttt{massimo.franceschet@uniud.it}}

\begin{abstract}
We analyse the large-scale structure of the journal citation network built from information contained in the Thomson-Reuters Journal Citation Reports. To this end, we take advantage of the network science paraphernalia and explore network properties like density, percolation robustness, average and largest node distances, reciprocity, incoming and outgoing degree distributions, as well as assortative mixing by node degrees. We discover that the journal citation network is a dense, robust, small, and reciprocal world. Furthermore, in and out node degree distributions display long-tails, with few vital journals and many trivial ones, and they are strongly positively correlated. 
\end{abstract}


\end{frontmatter}

\section{Introduction} \label{intro}

The present study is an interdisciplinary research integrating the fields of network science and bibliometrics. The field of \textit{network science} -- the holistic analysis of complex systems through the study of the structure of networks that wire their components -- exploded in the last decade, boosted by the availability of large databases on the topology of various real networks, mainly the Web and biological networks. The network science approach has been successfully applied to analyse disparate types of networks, including technological, information, social, and biological networks \citep{BE05,NBW06,N10}. Network analysis can be performed at different levels of aggregation:

\begin{itemize}
\item \textit{Node-level analysis}. At this level, the goal is to measure the importance or centrality of a node within the network. Centrality here is not an intrinsic and permanent feature of the node but, instead, it is an extrinsic and fleeting property that depends on the interactions of the node with the other nodes in the network. Typical node centrality measures include degree, eigenvector, closeness and betweenness centrality. 

\item \textit{Group-level analysis}. This investigation involves methods for defining and finding cohesive groups (clusters) of nodes in the network. The definition of cluster depends only on the topology of the network. Clusters are tightly knit sets of nodes with many edges inside the cluster and only a few edges between clusters. Two typical methods at this level of analysis are graph partitioning (where the number of clusters is fixed in advance) and community detection (in which the number of clusters is unspecified).

\item \textit{Network-level analysis}. The focus of this analysis is on properties of networks as a whole such as connectivity, mean and largest distances among nodes, distribution of node degrees, frequency of topological motifs, and assortative/disassortative mixing. It also includes the investigation on theoretical models explaining the generation of networks with certain features (e.g., random, small-world, and scale-free models). 
\end{itemize}
 
\textit{Bibliometrics} is an older field; it is a branch of information and library science that quantitatively investigates the process of publication of research achievements \citep{G55,P65}. Networks abound in bibliometrics; two important examples are citation networks of articles, journals or disciplines and collaboration networks of scholars. Other bibliometric networks are co-citation and co-reference networks of articles, journals or disciplines. 

Collaboration networks have been largely studied using the network science approach \citep{N04,BJNRSV02,G02,M04,RCCLP04,F11-JASIST}. Journal citation networks have been investigated mainly at node- and group-levels. The investigation at the node-level concerns the proposal of eigenvector-based centrality measures for journals \citep{PN76,BRS06,WBB10}, the clustering of journal bibliometric indicators, including centrality measures, on the basis of the statistical correlation among them \citep{L09,BSHC09}, as well as the use of betweenness centrality as an interdisciplinary indicator for journals \citep{LR11}. The group-level analysis of journal citation networks focuses on the detection, using different methods, of communities of journals, which correspond to fields of knowledge in the map of science \citep{L04,RB08,KB09,LMG10}. 

The investigation of journal citation networks at the network-level has been mainly focused on the study of the distribution of citations among papers and journals \citep{S92,R98,SSN08,RFC08}. The aim of the present investigation is to complement this investigation with additional large-scale structure properties of journal citation networks. More specifically, we focus on the following network properties: density of citation links, robustness with respect to the removal of nodes according to different percolation strategies, average and largest path lengths, topological motifs of reciprocity, incoming and outgoing degree distributions and their statistical correlations, as well as assortative mixing with respect to incoming and outgoing node degrees. 

\section{The large-scale structure of journal citation networks} \label{structure}

We considered all science and social science journals indexed in Thomson-Reuters Journal Citation Reports. We built a \textit{journal citation network} in which the nodes are the selected journals and there is a directed edge from node A to node B if journal A published  in 2008 a paper that cites a paper printed in journal B in the temporal window between 2003 and 2007. We only took into account the document types article and review. We considered the sub-network corresponding to the largest strongly connected component of the original network, which covers the large majority of the original network.\footnote{It is typical in network science to focus the analysis on the largest component when this is a \textit{giant} one, that is, when it includes the large majority of the nodes of the network \citep{N10}.} The resulting network is a directed unweighted graph with 6708 nodes and 1,315,238 edges between journals with the property that there exists a directed path between any pair of nodes. We loaded the network in the R environment for statistical computing \citep{R} and analysed the structure of the network using the  R package \textit{igraph} developed by G\'{a}bor Cs\'{a}rdi and Tam\'{a}s Nepusz. 

\begin{figure}[t]
\begin{center}
\includegraphics[scale=0.40, angle=-90]{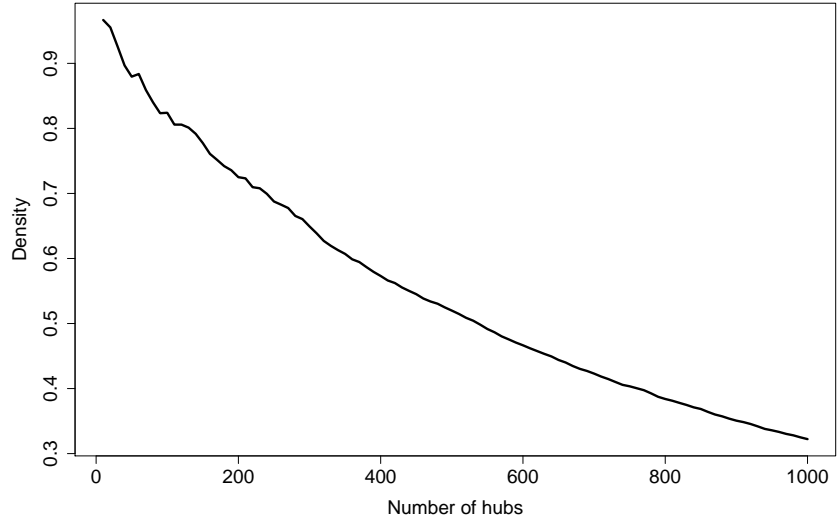}
\caption{The density of the network of top journals (journals with high total degree). The $x$ axis shows the number of top journals (up to 1000) and the $y$ axis gives the density of the network containing only these journals.}
\label{density}
\end{center}
\end{figure}

\begin{figure}[t]
\begin{center}
\includegraphics[scale=0.45, angle=-90]{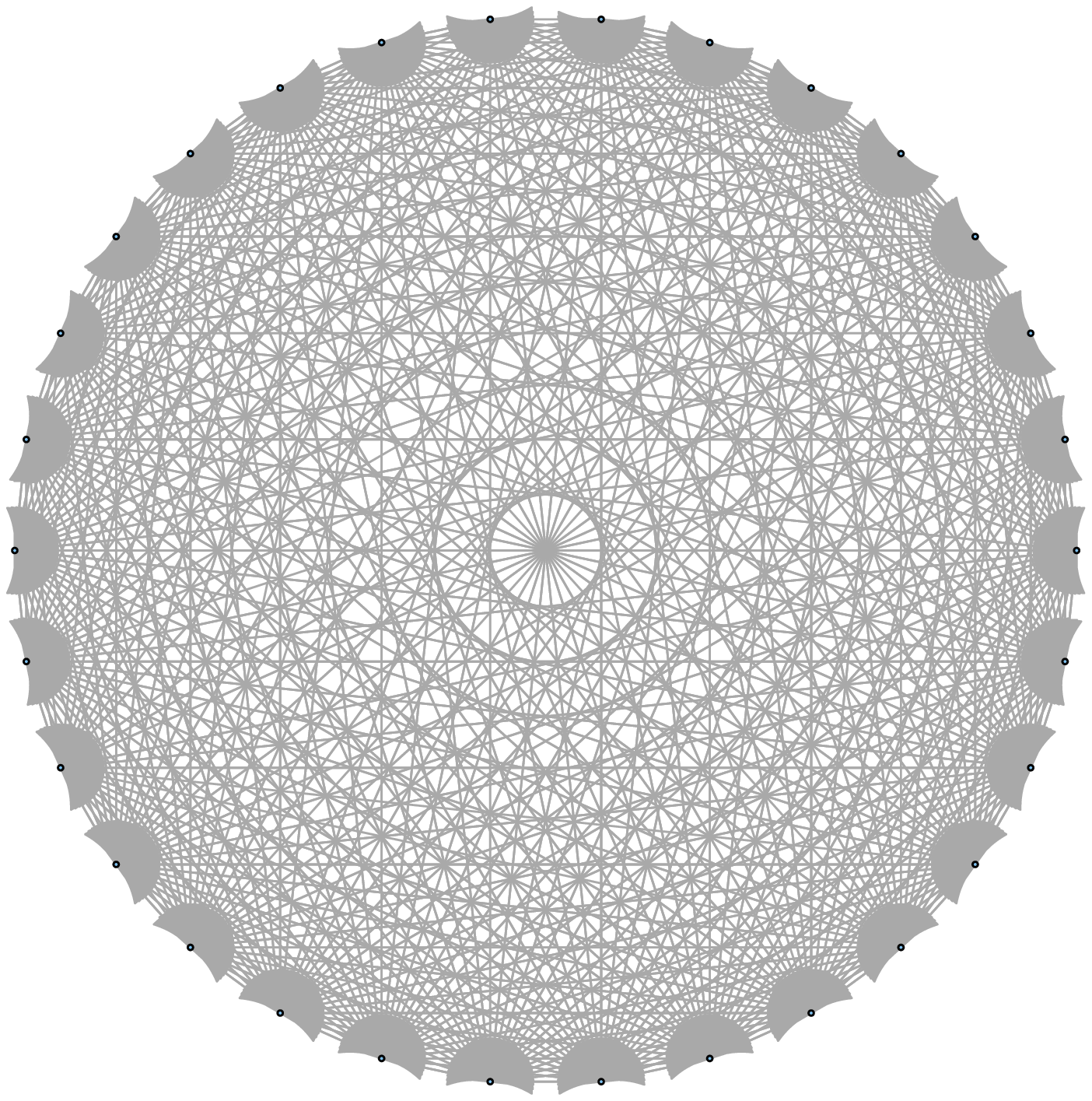}
\caption{The citation network of the top-30 journals. The (almost) complete graph resembles some Escher's works.}
\label{escher}
\end{center}
\end{figure}

The first network property that we analyse is density. Graph \textit{density} is the relative fraction of edges in the graph, that is the ratio of the actual number of edges and the maximum number of possible edges in the graph. The density of the journal citation graph is 3\%, meaning that the graph has 3 edges every 100 possible links between nodes. The density is much higher if we consider only top journals with large \textit{total degree}, where the total degree of a journal is the sum of the number of incoming edges (citing journals) and the number of outgoing edges (cited journals) of the journal in the network. We sorted the journals in decreasing order of total degree and we computed the density of the graph containing only an increasing share of top journals; the corresponding plot is depicted in Figure~\ref{density}. For instance, when only the top-30 journals are considered, the citation network, which is shown in Figure~\ref{escher}, is almost complete\footnote{A \textit{complete graph}, or \textit{clique}, is a graph with all possible edges.}, with a remarkably high density of 93\% (only 64 edges out of the 870 possible edges are missing). Notably, the density is relatively high (32\%) also for the network of top-1000 journals.

The journal citation graph is, by construction, \textit{strongly connected}. This means that there exists a directed path of citations between any pair of journals in the graph: a researcher can start reading any journal in any subject, e.g., tribology, and by following links of citations, they can reach any other journal in any other subject, e.g., mycology. Related to connectedness of a network is the concept of \textit{robustness}. Network robustness is typically investigated with a dynamic process called \textit{percolation}. The percolation process progressively removes nodes, as long as the edges connected to these nodes, from the network, and studies how the connectivity of the network changes. In particular, one wants to find the fraction of nodes to remove from the network in order to disintegrate its giant strongly connected component into small components. If such a fraction is relatively large, then the network is said to be robust to the process of percolation. To realize the percolation process, we progressively removed nodes from the collaboration network and, after each removal, we computed the relative size of the largest strongly connected component of the resulting sub-network. We tested the following node removal strategies \citep{N10}: 

\begin{enumerate}
\item \textit{degree-driven percolation}, in which the nodes are removed in decreasing order of node total degree (the sum of the in-degree and the out-degree of the node); 

\item \textit{eigenvector-driven percolation}, in which the nodes are removed in decreasing order of eigenvector centrality scores. A node has high eigenvector score if it is pointed to by nodes which, recursively, have high eigenvector scores;

\item \textit{closeness-driven percolation}, in which the nodes are removed in decreasing order of closeness centrality scores. A node has high closeness score if the mean distance from the node to all other nodes in the network is low;

\item \textit{betweenness-driven percolation}, in which the nodes are removed in decreasing order of betweenness centrality scores. A node has high betweenness score if the node lies on many geodesics (shortest paths) between other nodes in the network.

\end{enumerate}

\begin{figure}[t]
\begin{center}
\includegraphics[scale=0.40, angle=-90]{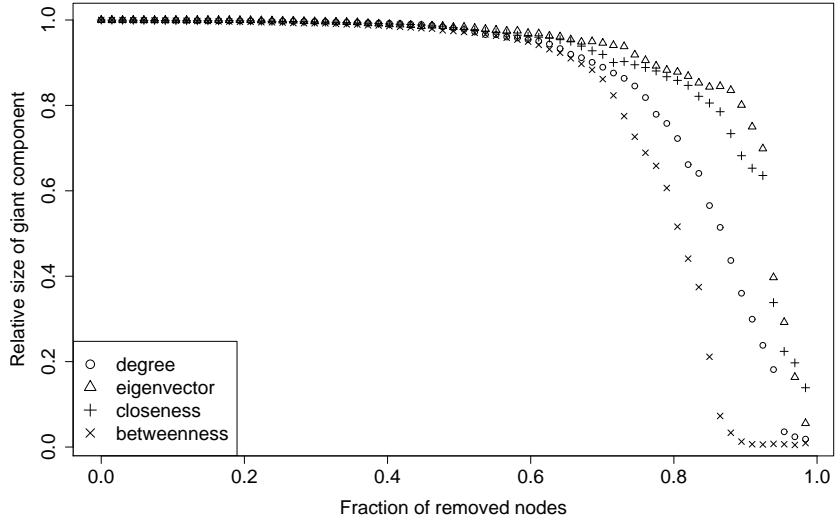}
\caption{Network robustness using the percolation process. The fraction of nodes removed from the network is plotted against the relative size of the largest strongly connected component. Four centrality strategies have been tested to remove the nodes in the percolation process.}
\label{robust}
\end{center}
\end{figure}

Figure \ref{robust} shows the outcomes of the described percolation process. A couple of observations emerge from the plot. First, the best percolation strategy is based on the removal of nodes in order of betweenness centrality. It dominates the total degree strategy, which is better than the closeness one. The least effective percolation strategy is the one based on eigenvector centrality.  Hence, when the objective is to dismantle the (strong) connectivity of the network, removing  `broker' nodes (nodes with high betweenness) is more effective than removing nodes with high total degree. Nodes with high betweenness score have been associated with interdisciplinary journals \citep{LR11}, while those with high total degree typically correspond to review journals. It follows that interdisciplinary journals are more responsible to keep the citation network strongly connected than review publication sources. Second, no percolation strategy is in fact really effective. Using the best percolation strategy (betweenness), 82\% of the nodes (almost the entirety of the graph) should be removed to reduce the largest strongly connected component below 50\% of the graph.  This is a distinct sign of the strong robustness of the journal citation network.

\begin{figure}[t]
\begin{center}
\includegraphics[scale=0.40, angle=-90]{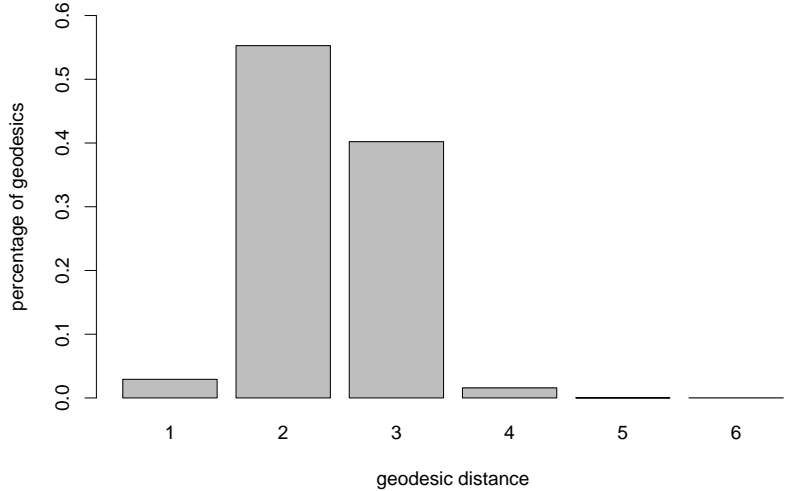}
\caption{Histogram of geodesic distances. For any given geodesic distance from 1 to the diameter of the network (6), a bar shows the percentage of geodesics having that distance.}
\label{geodesics}
\end{center}
\end{figure}

The fact that the journal citation network is strongly connected does not tell us anything about the lengths of paths in the network. For instance, compare a graph composed of a circle of nodes and a complete graph in which there is an edge connecting each pair of nodes. Both graphs are strongly connected, but the average distance between nodes in the first graph is of the order of the number of nodes, while in the second graph it is just 1. The \textit{geodesic distance} between two nodes in a graph is the number of edges of a shortest path (also known as geodesic) connecting the two nodes. We computed the geodesic distance for any pair of nodes in the graph. Figure \ref{geodesics} shows the geodesic distance histogram. The average geodesic distance is remarkably short: 2.4 edges. This means that given a random pair of journals, we can expect that in two or three citation hops we get from one journal to the other. The maximum geodesic distance, known as the \textit{diameter} of the network, is just 6 links (there are 20 paths with this largest length). We may conclude that the journal citation network is a \textit{small world} \citep{WS98}, in the sense that the average node distances are remarkably short (logarithmic) with respect to the number of nodes of the network. 

It is worth noticing that the average path length on co-reference networks have been recently proposed by \citet{RM10} in the context of indicators of interdisciplinarity. The authors investigate the interdisciplinary research in terms of two aspects: diversity (the number, balance and degree of difference between the bodies of knowledge integrated) and coherence (the extent that specific topics, concepts, tools and data used in a research process are related). In particular, they propose the mean path length defined on paper bibliographic coupling networks as a possible operationalization of the concept of network coherence. 

Real complex networks possess basic building blocks or \textit{motifs}: patterns of interconnections occurring in complex networks at numbers that are significantly higher than those in randomized networks \citep{MSIKCA02}. Such motifs have been found in diverse networks from biochemistry, neurobiology, ecology, and engineering. It is conjectured that these patterns play the role of functional circuit elements of the complex system underlying the network. The simplest motif that can be studied on a directed network is the loop of length two. On the journal citation network, this corresponds to a pair of journals that reciprocally cites themselves. This concept is known as \textit{reciprocity} in network science and it is operationalized by counting the relative frequency of edges that belong to a loop of length two in the network \citep{N10}. We computed reciprocity for the journal citation network and the result is 0.29; this means that 29\% of the times that a journal A cites another journal B we have that B cites back to A. This high percentage can be explained with the well-known phenomenon that journals can be partitioned into highly-connected clusters corresponding to disciplines and fields of them when journals are displayed on a citation map (see, for instance, \citet{RB08}).

\begin{figure}[t]
\begin{center}
\includegraphics[scale=0.40, angle=-90]{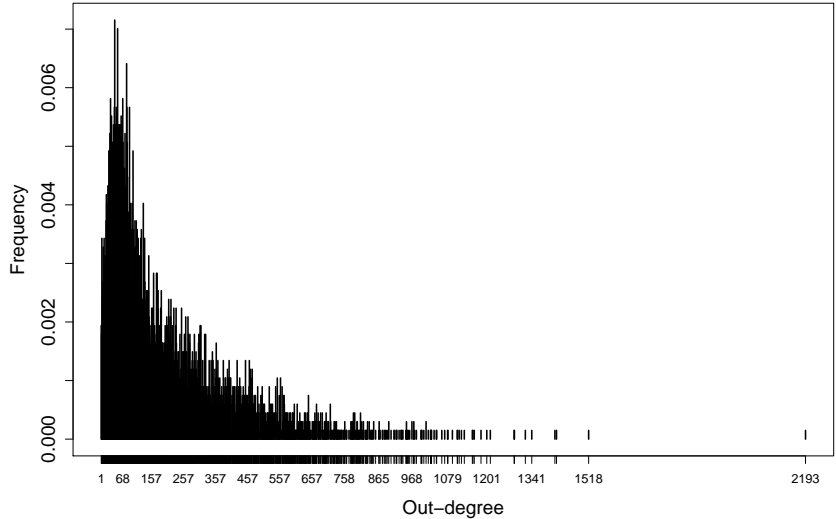}
\caption{The node out-degree (number of cited journals) distribution.}
\label{outdegree}
\end{center}
\end{figure}

\begin{figure}[t]
\begin{center}
\includegraphics[scale=0.40, angle=-90]{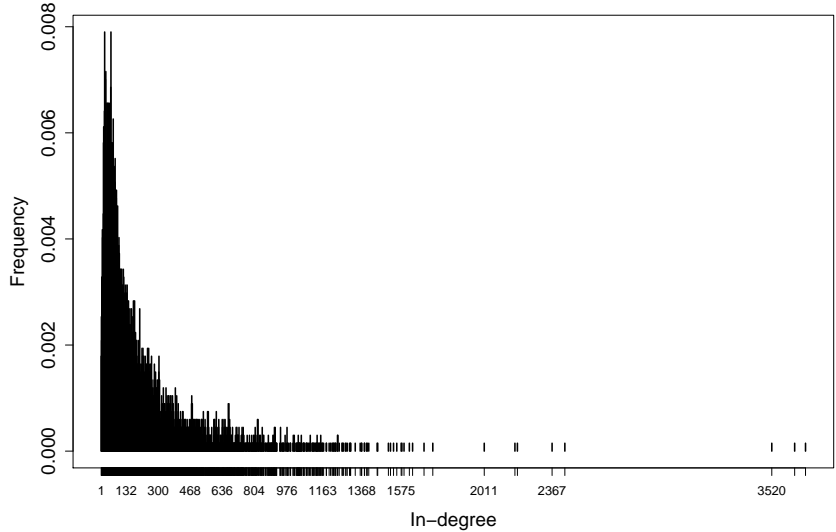}
\caption{The node in-degree (number of citing journals) distribution.}
\label{indegree}
\end{center}
\end{figure}

Finally, we investigate the node degree distributions of the journal citation network. Since the citation network is a directed graph, each journal has an \textit{out-degree} -- the number of distinct journals \textit{cited} by the journal, or the number of edges leaving the journal node --, and an \textit{in-degree} -- the number of distinct journals \textit{citing} the journal, or the number of edges arriving to the journal node. Figures \ref{outdegree} and \ref{indegree}, respectively, depict the out-degree and the in-degree distributions for nodes of the journal citation network. Both distributions have a clear long-tail: most of the journals cite and are cited by a relatively small number of other journals, but there is a significative number of \textit{hubs} -- journals that cite a large amount of other journals --, and \textit{authorities}, journals that are cited by a big number of other journals. The average degree for both distributions is 196.\footnote{This is the same number since each edge leaving a node is arriving at a node; this is also equal to the ratio of the number of edges and the number of nodes of the network.} The median out-degree is 126, with a maximum out-degree of 2193 (a third of the total number of journals) accomplished by journal PNAS. The median in-degree is 109, with a maximum in-degree of 3697 (more than half of the total number of journals) obtained by journal Science (PNAS is second with 3640). The in-degree distribution is more skewed and concentrated (skewness index is 3.5 and Gini concentration coefficient is 0.55) than the out-degree distribution (skewness index is 1.9 and Gini concentration coefficient is 0.49). Neither distribution follow a power-law, so the network cannot be regarded as a scale-free network \citep{BA99}.\footnote{We preliminarily observed a clear curvature of the complementary cumulative distribution function plotted on a double logarithmic scale; furthermore we performed \citet{CSN09} goodness-of-fit test for the power-law and log-normal models. Both tests excluded a (statistically significant) fit of the empirical distributions of degrees to the surveyed theoretical models (not even in a tail portion of the distribution for the power-law model).} 

Furthermore, the two distributions (in- and out-degree) are positively correlated (Spearman and Pearson correlation coefficients are 0.90 and 0.87, respectively): this means that there is a tendency for authority journals to be also hub journals and vice versa. This outcome is not crucially influenced by the size of the journal (in terms of number of published papers); indeed, the correlations of journal in-degree and out-degree with journal size are moderate (Spearman and Pearson correlation coefficients are 0.72 and 0.55, respectively). We also raised the following  questions: do authorities prefer to cite other authorities/hubs? Do hubs prefer to cite other hubs/authorities? In network science, \textit{assortative/disassortative mixing} is the tendency of nodes to connect to other nodes that are like/dislike them in some way \citep{N10}. The concept is implemented as a Pearson correlation coefficient over the investigated scalar characteristic (in- or out-degrees in our case) for nodes connected by an edge in the network. The correlation is positive and statistically significant in all four cases: 0.08 for authority/authority mixing, 0.14 for hub/hub mixing, 0.11 for authority/hub mixing, and 0.08 for hub/authority mixing. The low magnitude of the correlation coefficients is not surprising: most networks are naturally disassortative by degree because they are \textit{simple} graphs (at most one edge is possible between two nodes). Hence, a positive correlation in this case, although not large in magnitude, indicates a real assortativity by degree. In particular, networks  having a community structure override this natural bias and become assortative \citep{N10}.

\section{Conclusion}

We have analysed the journal citation network extracted from Thomson-Reuters Journal Citation Reports. Our conclusions are summarized in the following:

\begin{itemize}

\item the journal citation network has high reciprocity and positive assortativity by degree, which is coherent with a \textit{community structure}, in which there are tightly interconnected sets of journals that most likely represent entire disciplines or fields of them;

\item the journal citation network is a \textit{dense} and \textit{small} world. This means that, although the network is divided into closely integrated communities, there are quite intense inter-community flows of information (citations), and hence information can spread quickly over the whole academic community;

\item the journal citation network is highly \textit{robust}. These is good news for the whole academic community, since it means that there exists no restricted circle of influential journals that control the connectivity of the network and the diffusion of information on the whole academic community, although there are journals that are very influential within their local fields;

\item \textit{interdisciplinary journals} are more crucial than review sources for the connectivity of the network and for the diffusion of information over the academic community. The identification of interdisciplinary journals is a hot, partially open problem; interdisciplinarity is often perceived as a mark of good research, more successful in achieving breakthroughs and relevant outcomes \citep{RM10};

\item the degree distribution of the journal citation network shows a \textit{long tail} with many poorly endorsed journals an a significant few highly cited ones; the empirical distribution, however, does not match well the power-law model. To test the adherence to the power-law model we used the principled statistical framework developed by \citet{CSN09}. The same method is used by the developers to analyse a large number of real-world data sets from a range of different disciplines, each of which has been conjectured to follow a power law distribution in previous studies. Only two-third of them passed the test, and all of them showed the best adherence to the model when a (limited) suffix of the distribution is considered. 

\end{itemize}

\bigskip
\bigskip
\noindent
\textbf{Acknowledgements}

\medskip
\noindent I would like to thank Ludo Waltman (Centre for Science and Technology Studies, Leiden
University) for his assistance in the data collection.

\bibliographystyle{elsarticle-harv}

\begin{thebibliography}{31}
\expandafter\ifx\csname natexlab\endcsname\relax\def\natexlab#1{#1}\fi
\expandafter\ifx\csname url\endcsname\relax
  \def\url#1{\texttt{#1}}\fi
\expandafter\ifx\csname urlprefix\endcsname\relax\def\urlprefix{URL }\fi

\bibitem[{Barab\'{a}si and Albert(1999)}]{BA99}
Barab\'{a}si, A.-L., Albert, R., 1999. Emergence of scaling in random networks.
  Science 286, 509--512.

\bibitem[{Barab\'{a}si et~al.(2002)Barab\'{a}si, Jeong, N\'{e}da, Ravasz,
  Schubert, and Vicsek}]{BJNRSV02}
Barab\'{a}si, A.~L., Jeong, H., N\'{e}da, Z., Ravasz, E., Schubert, A., Vicsek,
  T., 2002. Evolution of the social network of scientific collaborations.
  Physica A: Statistical Mechanics and its Applications 311~(3-4), 590--614.

\bibitem[{Bollen et~al.(2009)Bollen, de~Sompel, Hagberg, and Chute}]{BSHC09}
Bollen, J., de~Sompel, H.~V., Hagberg, A., Chute, R., 2009. A principal
  component analysis of 39 scientific impact measures. PLoS ONE 4, e6022.

\bibitem[{Bollen et~al.(2006)Bollen, Rodriguez, and de~Sompel}]{BRS06}
Bollen, J., Rodriguez, M.~A., de~Sompel, H.~V., 2006. Journal status.
  Scientometrics 69~(3), 669--687.

\bibitem[{Brandes and Erlebach(2005)}]{BE05}
Brandes, U., Erlebach, T. (Eds.), 2005. Network Analysis: Methodological
  Foundations. Vol. 3418 of Lecture Notes in Computer Science. Springer.

\bibitem[{Clauset et~al.(2009)Clauset, Shalizi, and Newman}]{CSN09}
Clauset, A., Shalizi, C.~R., Newman, M. E.~J., 2009. Power-law distributions in
  empirical data. SIAM Review 51, 661--703.

\bibitem[{de~Solla~Price(1965)}]{P65}
de~Solla~Price, D., 1965. Networks of scientific papers. Science 149, 510--515.

\bibitem[{Franceschet(2011)}]{F11-JASIST}
Franceschet, M., 2011. Collaboration in computer science: a network science
  approach. Journal of the American Society for Information Science and
  Technology 62~(10), 1992--2012.

\bibitem[{Garfield(1955)}]{G55}
Garfield, E., 1955. Citation indexes to science: a new dimension in
  documentation through association of ideas. Science 122, 108--111.

\bibitem[{Grossman(2002)}]{G02}
Grossman, J.~W., 2002. The evolution of the mathematical research collaboration
  graph. Congressus Numerantium 158, 201--212.

\bibitem[{Klanans and Boyack(2009)}]{KB09}
Klanans, R., Boyack, K., 2009. Toward a consensus map of science. Journal of
  the American Society for Information Science and Technology 60~(3),
  455–476.

\bibitem[{Leydesdorff(2004)}]{L04}
Leydesdorff, L., 2004. Top-down decomposition of the {Journal Citation Report}
  of the {Social Science Citation Index}: Graph- and factor-analytical
  approaches. Scientometrics 60~(2), 159--180.

\bibitem[{Leydesdorff(2009)}]{L09}
Leydesdorff, L., 2009. How are new citation-based journal indicators adding to
  the bibliometric toolbox? Journal of the American Society for Information
  Science and Technology 60~(7), 1327--1336.

\bibitem[{Leydesdorff et~al.(2010)Leydesdorff, de~Moya-Aneg\'{o}n, and
  Guerrero-Bote}]{LMG10}
Leydesdorff, L., de~Moya-Aneg\'{o}n, F., Guerrero-Bote, V.~P., 2010. Journal
  maps on the basis of {Scopus} data: A comparison with the {Journal Citation
  Reports of the ISI}. Journal of the American Society for Information Science
  and Technology 61~(2), 352--369.

\bibitem[{Leydesdorff and Rafols(2011)}]{LR11}
Leydesdorff, L., Rafols, I., 2011. Indicators of the interdisciplinarity of
  journals: Diversity, centrality, and citations. Journal of Informetrics 5,
  87--100.

\bibitem[{Milo et~al.(2002)Milo, Shen-Orr, Itzkovitz, Kashtan, Chklovskii, and
  Alon}]{MSIKCA02}
Milo, R., Shen-Orr, S., Itzkovitz, S., Kashtan, N., Chklovskii, D., Alon, U.,
  2002. Network motifs: Simple building blocks of complex networks. Science
  298~(5594), 824--827.

\bibitem[{Moody(2004)}]{M04}
Moody, J., 2004. The structure of a social science collaboration network:
  {D}isciplinary cohesion from 1963 to 1999. American Sociological Review
  69~(2), 213--238.

\bibitem[{Newman(2004)}]{N04}
Newman, M. E.~J., 2004. Coauthorship networks and patterns of scientific
  collaboration. Proceedings of the National Academy of Sciences of the United
  States of America 101, 5200--5205.

\bibitem[{Newman(2010)}]{N10}
Newman, M. E.~J., 2010. Networks: An introduction. Oxford University Press.

\bibitem[{Newman et~al.(2006)Newman, Barab{\'a}si, and Watts}]{NBW06}
Newman, M. E.~J., Barab{\'a}si, A.-L., Watts, D.~J., 2006. The Structure and
  Dynamics of Networks. Princeton University Press.

\bibitem[{Pinski and Narin(1976)}]{PN76}
Pinski, G., Narin, F., 1976. Citation influence for journal aggregates of
  scientific publications: Theory, with application to the literature of
  physics. Information Processing \& Management 12~(5), 297 -- 312.

\bibitem[{{R Development Core Team}(2008)}]{R}
{R Development Core Team}, 2008. R: A Language and Environment for Statistical
  Computing. R Foundation for Statistical Computing, Vienna, Austria, {ISBN}
  3-900051-07-0.
\newline\urlprefix\url{http://www.R-project.org}

\bibitem[{Radicchi et~al.(2004)Radicchi, Castellano, Cecconi, Loreto, and
  Parisi}]{RCCLP04}
Radicchi, F., Castellano, C., Cecconi, F., Loreto, V., Parisi, D., 2004.
  Defining and identifying communities in networks. Proceedings of the National
  Academy of Sciences of the United States of America 101~(9), 2658--2663.

\bibitem[{Radicchi et~al.(2008)Radicchi, Fortunato, and Castellano}]{RFC08}
Radicchi, F., Fortunato, S., Castellano, C., 2008. Universality of citation
  distributions: Toward an objective measure of scientific impact. Proceedings
  of the National Academy of Sciences of the United States of America 105~(45),
  17268--17272.

\bibitem[{Rafols and Meyer(2010)}]{RM10}
Rafols, I., Meyer, M., 2010. Diversity and network coherence as indicators of
  interdisciplinarity: Case studies in bionanoscience. Scientometrics 82,
  263--287.

\bibitem[{Redner(1998)}]{R98}
Redner, S., 1998. How popular is your paper? {An} empirical study of the
  citation distribution. The European Physical Journal B 4, 131--134.

\bibitem[{Rosvall and Bergstrom(2008)}]{RB08}
Rosvall, M., Bergstrom, C.~T., 2008. Maps of random walks on complex networks
  reveal community structure. Proceedings of the National Academy of Sciences
  of the United States of America 105, 1118--1123.

\bibitem[{Seglen(1992)}]{S92}
Seglen, P.~O., 1992. The skewness of science. Journal of the American Society
  for Information Science 43~(9), 628--638.

\bibitem[{Stringer et~al.(2008)Stringer, Sales-Pardo, and Amaral}]{SSN08}
Stringer, M.~J., Sales-Pardo, M., Amaral, L. A.~N., 2008. Effectiveness of
  journal ranking schemes as a tool for locating information. PLoS ONE 3~(2),
  e1683.

\bibitem[{Watts and Strogatz(1998)}]{WS98}
Watts, D.~J., Strogatz, S.~H., 1998. Collective dynamics of `small-world'
  networks. Nature 393, 440--442.

\bibitem[{West et~al.(2010)West, Bergstrom, and Bergstrom}]{WBB10}
West, J.~D., Bergstrom, T.~C., Bergstrom, C.~T., 2010. The {Eigenfactor}
  metrics: A network approach to assessing scholarly journals. College and
  Research Libraries 71, 236--244.

\end{thebibliography}

\end{document}